\documentclass[aps,prl,nofootinbib,preprintnumbers,reprint]{revtex4-2} 
\usepackage{amsmath,amssymb}
\usepackage{graphicx}  
\usepackage{bbold}
\usepackage{slashed}
\newcommand{\bL}{\begin{Large}}
\newcommand{\eL}{\end{Large}}

\newcommand{\bea}{\begin{eqnarray}}
\newcommand{\eea}{\end{eqnarray}}
\newcommand{\be}{\begin{equation}}
\newcommand{\ee}{\end{equation}}

\usepackage[colorinlistoftodos]{todonotes}
\usepackage{appendix}

\DeclareGraphicsExtensions{.pdf,.png,.jpg,.eps}
\graphicspath{ {./}{./figs/} }

\begin{document}

\preprint{\rightline{KCL-PH-TH/2025-{\bf 15}}}

\title{Superradiant Axionic Black-Hole Clouds as Seeds for Graviton  Squeezing}

\author{Panagiotis Dorlis$^a$}
\author{N. E. Mavromatos$^{a,b}$}
\author{Sarben Sarkar$^b$}
\author{Sotirios-Neilos Vlachos$^a$}
\medskip
\affiliation{$^a$Physics Division, School of Applied Mathematical and Physical Sciences, National Technical University of Athens, Zografou Campus, Athens 157 80, Greece}
\affiliation{$^b$Theoretical Particle Physics and Cosmology Group, Department of Physics, King's College London, London, WC2R 2LS, UK}

\date{\today}

\begin{abstract}
It is shown that both, standard general relativity (GR)  and Chern-Simons (CS)  gravity, the latter containing chiral gravitational anomaly terms, seed the production of pairs of entangled gravitons in a multi-mode squeezed state.  This  involves the interaction of gravitons with the axionic cloud surrounding a superradiant Kerr (rotating) black hole background. The order of magnitude of the squeezing effect, specifically the number of graviton excitations in the squeezed vacuum, is estimated in the non-relativistic limit,  relevant for  the superradiance process. It is found analytically that the  squeezing from the  GR process of annihilation of two axions into two gravitons, dominates, by many orders of magnitude,  that  coming from the axion decay into two gravitons, induced by the higher-derivative CS term. 
It is also shown that significant squeezing effects are produced in the case of long-lived axionic clouds, whose lifetimes are much longer than the timescale for which superradiance is effective. A brief discussion on current exclusion (for the first time) of very-long axion-cloud lifetimes, through comparison of our results with current LIGO data, as well as potential detection of such effects in future interferometers is also given. 

\end{abstract}

\maketitle

{\it \textbf{Introduction}}. It is well known that a mathematically and physically consistent theory combining both quantum theory and general relativity  still eludes us. However, there is no consensus on whether the gravitational
field is quantized  just like the other fundamental fields of nature \cite{str,str1,str2,loop,loop2,Verlinde,Verlinde2,as,as2}. In contrast to the detection of a single photon, it is a near-impossible task~\cite{Dyson:2013hbl,singlegrav,Carney:2023nzz} to detect a single graviton. However, collective states of gravitons might be detectable. GR can be consistently treated as an \textit{effective field theory} (EFT) \cite{Donoghue:1994dn, Donoghue:2022eay}, within the framework of the background-field method \cite{DeWitt:1967ub,tHooft:1974toh}. The validity of the EFT requires energies below the ultraviolet (UV) cut-off, whose upper limit is given by the (reduced) Planck mass scale, $M_{\rm Pl}=\kappa^{-1}= 2.435 \times 10^{18}$~GeV. This defines the \textit{perturbative quantum gravity} (PQG) approach, in which one considers a fixed classical background and small  quantum  perturbations about it. In  the PQG  regime, non-renormalizabilty issues are avoided and one  can  make  potentially  observable predictions. These include properties of  gravitational waves (GWs), which were observed for the first time in 2015~\cite{LIGOScientific:2016aoc}. The future prospects for optimizing both Earth-based and space-based GW detectors  promise  a new era of GW astronomy \cite{Bailes:2021tot}.  A  new cross-fertilization with 
ideas from quantum optics~\cite{PhysRevA.31.2409,Wu:1986zz,Boyd_optics,Scully_Zubairy_1997,Agarwal_2012,Lerch_2013}  may lead to the detection of the quantum nature of gravity in GWs.  

In  future  GW interferometric detectors~\cite{Amelino-Camelia:1998mjq,Amelino-Camelia:1999vks}, quantum fluctuations of the gravitational field might be detectable as an appropriate kind of {\it{quantum}} noise~\cite{Parikh:2020kfh}, depending on the state of gravitons. Squeezed states, which are states with no classical analogue, may  lead to  detectable signatures  for  a large enough squeezing parameter.  For example , the non-classicality in GWs can be explored by employing a Hanbury-Brown-Twiss (HBT) interferometer \cite{Kanno:2018cuk, Kanno:2019gqw},  which  can reveal sub-Poissonian statistics for the number of gravitons in squeezed coherent states.  However HBT analysis is purely theoretical, with no experimental implementation so far, since it relies on delicate intensity-correlation techniques related to indirect graviton counting.  Another approach uses the electromagnetic field cavity as an optomechanical GW detector \cite{Guerreiro:2019vbq,Coradeschi:2021szx}, where quantum properties of the graviton states can be revealed by statistical measurements.  For a single-mode squeezed vacuum state to have an observational signature, a larger than one squeezing parameter \cite{Scully_Zubairy_1997} is required. In~~\cite{Coradeschi:2021szx} the authors considered 
very large squeezing parameters of order $r\sim78$, corresponding to an excited vacuum containing $\langle N\rangle =\sinh^2 r\sim 10^{67}$ gravitons, as examples of potentially detectable graviton squeezing in future GW detectors. Nonetheless, the non-observation of squeezed GW states by the LIGO/Virgo GW interferometer~\cite{LIGOScientific:2016aoc,McCuller:2021mbn} imposes stringent upper bounds on the GW-squeezing parameter $r < 41$~\cite{Hertzberg:2021rbl}. The allowed region of $r$ still leaves plenty of room for sufficiently large squeezing parameters, and, as we shall see, it is compatible with our mutli-mode-squeezed-graviton considerations in this work.

Squeezed graviton states arise naturally during inflation. At cosmological scales, the expansion of the universe produces pairs of particles, forming a two-mode squeezed state with opposite momenta \cite{Mukhanov:2007zz,Kanno:2018cuk,Kanno:2019gqw}. 
The inflationary vacuum appears as a squeezed vacuum from the viewpoint of the radiation-era, via appropriate Bogoliubov transformations. From a non-cosmological perspective, astrophysical sources of non-classical GWs are crucial, though rarely discussed in the literature~\cite{Guerreiro_astro};  such sources could potentially increase the abundance of observational sources beyond those predicted by inflation.  Exploring astrophysical  frameworks  for producing  sufficient  ``squeezing'' lies at the heart of this letter. 
\begin{figure}[ht]
    \centering
    \includegraphics[width=0.9\linewidth]{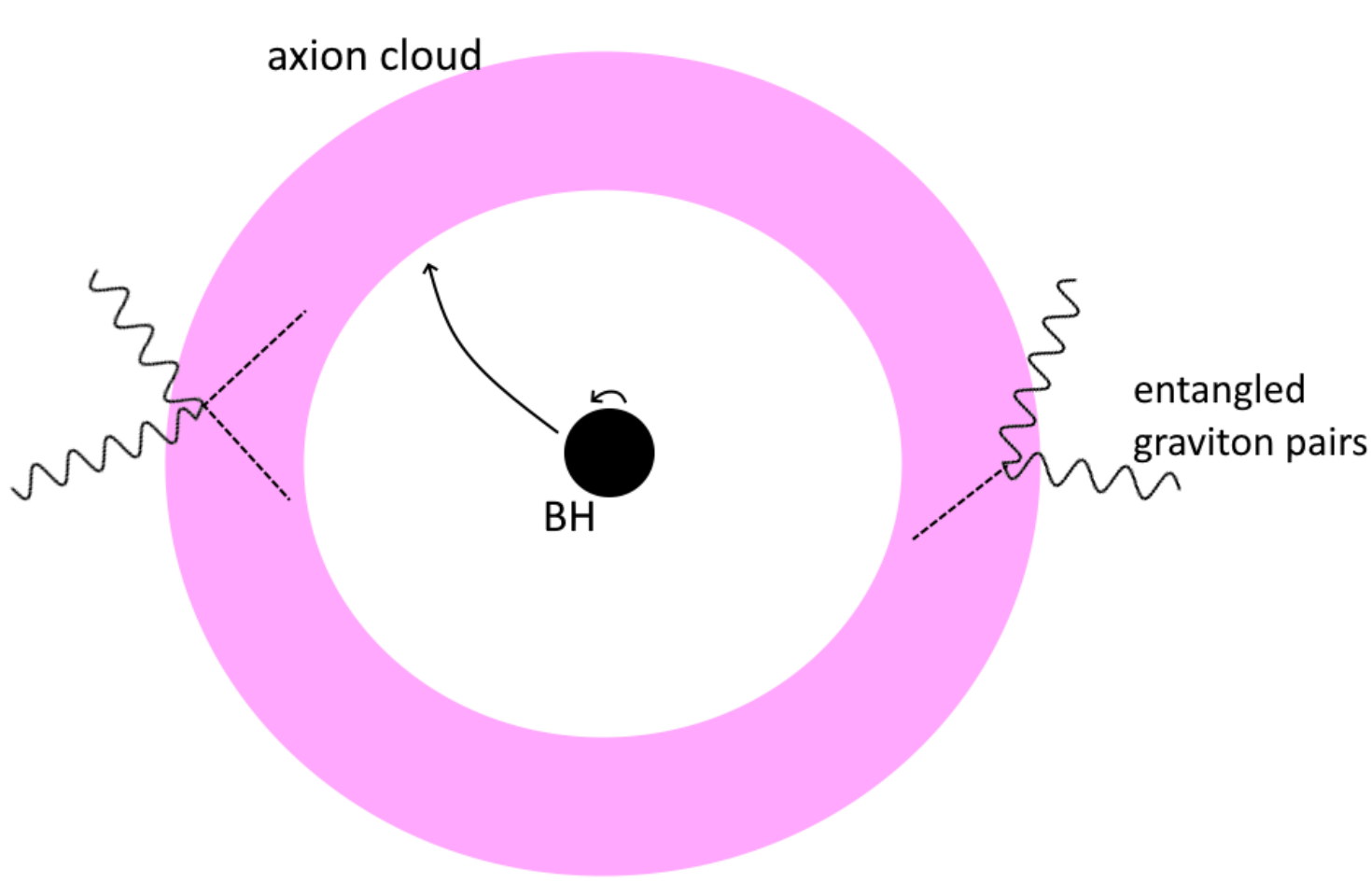}
    \caption{Superradiant axionic cloud: non-linear axion-graviton interactions producing entangled graviton pairs.  }
    \label{axion_cloud}
\end{figure}

In quantum optics, non-classical states of photons are produced by non-linear interactions in the presence of a medium. Spontaneous parametric down-conversion (SPDC) \cite{PhysRevA.31.2409, Wu:1986zz}, splits one high-energy (``pump’’) photon into two lower-energy entangled photons \cite{SPDC_multimode}. On the other hand, spontaneous four-wave mixing (SFWM)~\cite{SFWM1,SFWM2,SFWM_3} requires two ``pump’’ photons interacting inside a nonlinear medium, resulting in the production of entangled photon pairs. This leads to a quadratic dependence on the pump field, in contrast to the linear dependence in SPDC.

In this letter, we propose that gravitons in a multi-mode squeezed state are emitted due to the non-linear axion-graviton interaction.
Superradiant instability leads to the formation of an axionic cloud-condensate around a rotating BH, the so called ``gravitational atom" \cite{arvanitaki,arvanitaki2,arvanitaki3,arvanitaki4,BritoCardoso,VicenteCardoso} ({\it c.f.} figure \ref{axion_cloud}). The dominant squeezing process is that of the annihilation of two axions into a pair of entangled gravitons, the gravitational analogue of SFWM. 
This arises  from  the kinetic axion term in the GR lagrangian, which contains a coupling of two axions  to two gravitons. The gravitational Chern-Simons (CS) anomaly~\cite{Jackiw:2003pm,Duncan:1992vz,Alexander:2009tp} around a rotating BH, induces a decay process of a single axion into two gravitons  and provides  the gravitational analogue of SPDC. Because of  the large occupation number of axions in the condensate, such  decay amplitudes  are significantly enhanced  over,  amplitudes for single or two-axion Fock states;  the axion classical background mimics the coherent pump field in SPDC/SFWM. 

The axionic fields in the BH cloud  can  have  diverse origins.  Our  axions are either string-compactification axions or string-model-independent axions (Kalb-Ramond (KR) axions) ~\cite{Svrcek:2006yi,Duncan:1992vz} or are equivalent to the totally antisymmetric components of torsion in Einstein-Cartan models~\cite{Duncan:1992vz,iorio} for contorted spacetime geometries, provided mechanisms for making all such axions massive are in place.

{\it \textbf{Superradiance}}. It is known that a rotating black hole (BH) has a superradiant instability, where bosonic particles are abundantly produced \cite{BritoCardoso}. Upon extracting rotational energy from the BH, these particles exist in quasibound states around the BH  interpreted as  a condensate-like distribution. In a background of a Kerr BH~\cite{Kerr:1963ud} of mass $\mathcal{M}$, angular momentum  $\mathcal{J}_H$ and spin parameter $ \alpha = \mathcal{J}_H/\mathcal{M}$, the Klein-Gordon equation (KG)  for a massive (pseudo)scalar field, of mass $\mu_b$, admits quasibound states, which are labeled by integer numbers $\left(n,l,m\right)$. We shall denote as $r_{\pm}$ the outer ($+$) and inner ($-$) event horizons, respectively. 
We shall work in the non-relativistic regime \cite{Detweiller}, that is, when the Compton wavelength of the axion $\lambda_C\sim1/\mu_b$ is much larger than the black hole radius and the BH-axion binding energy is small compared to the axion rest mass. In this regime, the angular part of the solution of KG is described by spherical harmonics, while the effective radial equation reduces to a Coulomb-like problem, with solutions resembling the wavefunctions of the hydrogen atom \cite{Chandrasekhar1998-zk}. 
The imaginary part of the frequency \(\omega_I\) drives the superradiant instability with a growth $\sim e^{\omega_I t}$, provided the real part of the frequency, $
\omega$, satisfies  the superradiant condition~\cite{Bekenstein:1973mi} $\omega<m  \Omega_{H}$, where $\Omega_H$ is the angular velocity of the event horizon. For the real part of the frequency, we have~\cite{BritoCardoso}:
\[
\omega \approx \mu_b\left(1 - \frac{a_{\mu}^2}{2n^2}\right) \quad, \quad  a_\mu \equiv G \mathcal{M}\mu_b \ ,
\]
with $a_\mu$ the dimensionless coupling of the gravitational atom. In the non-relativistic regime~\cite{Detweiller}, we have  $a_\mu \ll 1$. In this approach, the classical axion field is given by:
\begin{equation}
    b(t,r,\theta,\phi)=\sum_{nlm}e^{-i \omega_n t} \sqrt{\frac{N_{nlm}}{2\mu_b}}\Psi_{nlm} + \text{c.c} \ ,
\end{equation}
where $N_{nlm}$ is chosen to be the number of axions in the respective state, with $\Psi_{nlm}(\vec{x})$ obeying the  normalization $\int d^3x \vert \Psi_{nlm}\vert^2
=1$. As long as the superradiance condition $\omega<m \Omega_H$ is satisfied, the axionic-cloud will grow at a rate faster than the evolution timescale of the BH  (see discussion below).
The most dominant mode corresponds to the ``$2p$-axion state" $(n=2, \ l=m=1)$. Regarding timescales relevant to superradiance, there are three main processes that govern the evolution: the superradiant growth of the instability, the accretion of black holes and GW emission  \cite{BritoScales1,BritoScales2}. The instability time $\tau_s$ of superradiance for the $2p$-state scales as \cite{Detweiller}:
\begin{equation}
\label{timescale_superradiance}
\tau_s=\frac{1}{\omega_I (2p)}=24\ \mu_b^{-1}\left(\frac{\alpha}{G\mathcal{M}}\right)^{-1}\, a^{ -8}_\mu \ ,
\end{equation}
where we consider highly rotating BHs, $\alpha/(G \mathcal{M}) \sim \mathcal{O}(1)$ and $a_\mu=0.1$, consistent with the non-relativistic approximation \cite{BritoCardoso,Detweiller}. Note that  $\tau_s$ is a separate timescale from the oscillation period of the cloud $\tau\sim\mu_b^{-1}$, and therefore, a quasi-stationary approximation is valid. The evolution of the superradiant instability unfolds in two stages: first, the axionic condensate evolves due to superradiance around the BH background, on a timescale given by $\tau_s$ \eqref{timescale_superradiance}. Then, the cloud emits energy via GWs, a process characterized by the relevant timescale $\tau_{GW}$. There is an important separation of these two scales ($\tau_{GW} \gg \tau_s$), ensuring two basic physical features; i) the cloud can evolve to saturation without being depleted through GWs emission and ii) the axionic-condensate is very long lived, as supported by various 
perturbative calculations of GW emission by the cloud~\cite{BritoScales1,BritoScales2,Yoshino_2014_scales,Brito_2017_scales}.  Since the lifetime $T$ of the cloud has to take into account subprocesses with widely separated time scales, it is considered to be a phenomenological parameter.

The maximum number of axions occupying the dominant mode is given by \cite{arvanitaki3,Arvanitaki_2017_Number_axions,Bernal_2022}:
\begin{equation}
    N_{2p}^{max}\sim 10^{74}\left(\frac{\Delta \alpha_\star}{0.1}\right)\left(\frac{\mathcal{M}}{M_{\odot}}\right)^{2} \ ,
    \label{number_of_axions}
\end{equation}
where $\alpha_\star=\alpha/(G\mathcal{M})$ and $\Delta \alpha_\star=\mathcal{O}(0.1)$ denotes the difference between the initial and final BH spins. The radii of the ``$2p$-axionic cloud" is given by $\langle r_{c} \rangle = 5 r_0\,$
with variance $\Delta r_{c} = \sqrt{5}  r_0$, where $r_0=(a_\mu\mu_b)^{-1}$
denotes the ``gravitational Bohr radius". In the non-relativistic limit $\left(a_\mu \ll 1\right)$, $r_c$ is much larger than the dimensions of the black hole $r_+ \sim G \mathcal{M}$. 
Therefore, we may proceed by ignoring the curvature effects and quantizing GWs in a flat spacetime background.

{\it \textbf{Axion - Graviton Interactions}}. 
The relevant gravitational interactions,  showing  the role of axions as seeds of the superradiant instability, are the GR-induced minimal coupling, associated with the axion Lagrangian, and the gravitational CS topological term (Hirzebruch signature)~\cite{Jackiw:2003pm,Duncan:1992vz,Alexander:2009tp},
\begin{align}\label{Rcs}
    S_{CS}  =-\frac{A}{2}\int d^{4}x\sqrt{-g}\ \ b\ \, R_{\mu\nu\rho\sigma}\widetilde{R}^{\nu\mu\rho\sigma} \  , 
\end{align}
where $A$ denotes the coupling constant, and $\widetilde{R}_{\mu\nu\rho\sigma}=\frac{1}{2}R_{\mu\nu\alpha\beta}\ \epsilon^{\alpha\beta}\!_{\rho\sigma}\ $ is the dual of the Riemann tensor, with  $\epsilon_{\mu\nu\rho\sigma}$ the covariant Levi-Civita tensor. 

In the CS gravitational theory \cite{Jackiw:2003pm}, black holes exhibit axionic hair \cite{Yagi:2012ya,Duncan:1992vz,Chatzifotis:2022mob}, which is a  stationary configuration that solves the gravitational equations of motion. However, superradiance in the presence of the anomaly \eqref{Rcs} corresponds to non-stationary solutions of the (pseudo)scalar Klein-Gordon equation in the Kerr background. In both cases, the effect of higher-curvature couplings is highly suppressed by the ratio of the Planck mass over the BH mass; the effect is negligible, unless we are dealing with {\it{extremely small}} BHs, with masses comparable to the Planck mass. Therefore, in the decoupling limit \cite{Teukolsky}, the process of superradiance can be considered identical to that in the GR case \cite{Richards:2023xsr}.

Upon assuming a flat background spacetime $\eta_{\mu\nu}$, in the weak field expansion, $\eta_{\mu\nu} \to \eta_{\mu\nu}+ \kappa h_{\mu\nu}$, $\kappa h_{\mu\nu}\ll1$, the metric perturbation in the transverse and traceless (TT) gauge yields the following axion-graviton interactions:
\begin{align}
    \label{S1}
    S^{(1)}&= \frac{\kappa}{2}\int d^4x \ h_{ij}T^{ij}  \ ,  \\ 
    \label{S2}
    S^{(2)}&=  - \frac{\kappa^2}{2} \int d^4x \ h_{im}h^{m}_{\ j} \ \partial^i b\,\partial^j b \ ,  \\
    \label{S2CS}
     \nonumber S^{(2)}_{CS} & = -A\kappa^2\int d^4x \ b(x) \epsilon_{ijk} \ \Bigg( \partial_l\partial^k  h^j_m  \partial^m\dot{h}^{li} \\ & +\ddot{h}^{li}\partial^k\dot{h}^j_l  
    -\partial_m\partial^kh^j_l\partial^m\dot{h}^{li}  \Bigg)  
\end{align}
 where $T_{ij}$ is the stress-momentum tensor (and Latin indices are 3-space ones, $i, j, \dots = 1,2,3)$. 
Expanding the perturbations in Fourier space, one can quantize the system by placing it in an initially finite volume $V$:
\begin{equation}
    \hat{h}_{ij}(t,\vec{x})=M_{\rm Pl}\sum_{\vec{k},\lambda} f_k\left[   e^{(\lambda)}_{ij}(\vec{k}) \hat{\alpha}^\dagger_{\lambda,\vec{k}}\ e^{-ik\cdot x} +h.c  .  \right] \ ,
    \label{mode_expansion_gravitons_}
\end{equation}
where $\hat{\alpha}_{\lambda,\vec{k}},\hat{\alpha}^{\dagger}_{\lambda,\vec{k}}$ denotes the  dimensionless annihilation/creation operators obeying the usual commutation relations,
$ [\hat{\alpha}_{\lambda,\vec{k}}  ,  \hat{\alpha}^\dagger_{\lambda^\prime , \vec{k}^\prime}] = \delta_{\lambda\lambda^\prime} \delta_{\vec{k}\vec{k}^\prime}$, where $\lambda=L,R$ denotes left (L) and right (R) polarisation respectively,  $f_k$ is the (dimensionless) single graviton  strain, $f_k\equiv \kappa(2V\Omega_k)^{-\frac{1}{2}}$, while $\Omega_{k}\equiv k$ is the frequency of GW with momentum $\vec k$. The polarisation tensors obey the normalization
$e^{*(\lambda)}_{ij} e_{(\lambda^\prime)}^{ij}=2\delta_{\lambda\lambda^\prime}$. The interaction linear in the graviton \eqref{S1} is responsible for the production of graviton coherent states~\cite{Skagerstam:2018jkw}. The remaining interactions correspond to processes including two gravitons; the first one \eqref{S2}, stemming from GR, describes two axions annihilating into two gravitons, while the second one \eqref{S2CS}, associated with the anomaly, corresponds to an axion decay into two gravitons. 

Both interactions are capable of squeezing the graviton states. These phenomena are analogous to the quantum optics processes of SFWM and SPDC, respectively. The pertinent interaction Hamiltonian  has the general structure: 
\begin{equation}
\label{Hamiltonian_general_structure}
\begin{aligned}
    \hat{H}_{\text{int}} =\sum_{I,J}e^{i(\Omega_k+\Omega_{k^\prime}-E)t}  \mathcal{F}_{IJ} \hat{\alpha}^\dagger_I\hat{\alpha}^\dagger_J+  h.c  +  \dots 
    \end{aligned}
\end{equation}
where the index $I=(\lambda,\vec{k})$ denotes the graviton states, while $E$ denotes the available axion energy for the production of gravitons in each process. The $\dots$ in \eqref{Hamiltonian_general_structure} correspond to terms with mixed creation and annihilation operators for the graviton. Such interactions are ignored in the rotating wave approximation (RWA) \cite{WallsMilburn2008,PhysRevA.31.2409,Wu:1986zz,SFWM1,SFWM2,SFWM_3}, since, for a long enough interaction time, these terms oscillate rapidly and are thus subdominant. 

The evolution operator takes the form of a multimode squeezing operator \cite{PhysRevA.47.733}, 
\begin{equation}
\label{Evolution_Operator}
    \hat{S}=\exp\left[ \frac{1}{2}\sum_{I,J}\mathcal{G}_{IJ}\ \hat{\alpha}^\dagger_I\hat{\alpha}^\dagger_J -h.c.    \right]
\end{equation}
where, 
\begin{equation}
\label{G_IJ}
    \mathcal{G}_{IJ}=-2\ i\ \mathcal{F}_{IJ} \  T \ \text{sinc}\left[\left(\Omega_k + \Omega_{k^\prime} - E \right) \frac{T}{2}\right] 
\end{equation}
resembles the multi-mode squeezing parameter, with $T$ denoting the lifetime of the classical (coherent) source that drives the process; in our case, the lifetime of the axionic condensate (cloud) around the BH. We note that:  $$T\text{sinc}\left[\left(\Omega_k + \Omega_{k^\prime} - E \right) (T/2)\right]\stackrel{T\to\infty}{=}\delta\left(\Omega_k + \Omega_{k^{\prime}} - E\right),$$ implying energy conservation $\Omega_k + \Omega_{k^{\prime}} \approx E$. Note that, in the definition of the evolution operator, the time ordering of Dyson's formula has been omitted under the RWA. This becomes more evident when one considers the Magnus expansion for the evolution operator \cite{magnus1, magnus2}, which is an expansion of the exponent in terms of commutators of the Hamiltonian at different times. 

\begin{figure}[ht]
    \centering
    \includegraphics[width=0.9\linewidth]{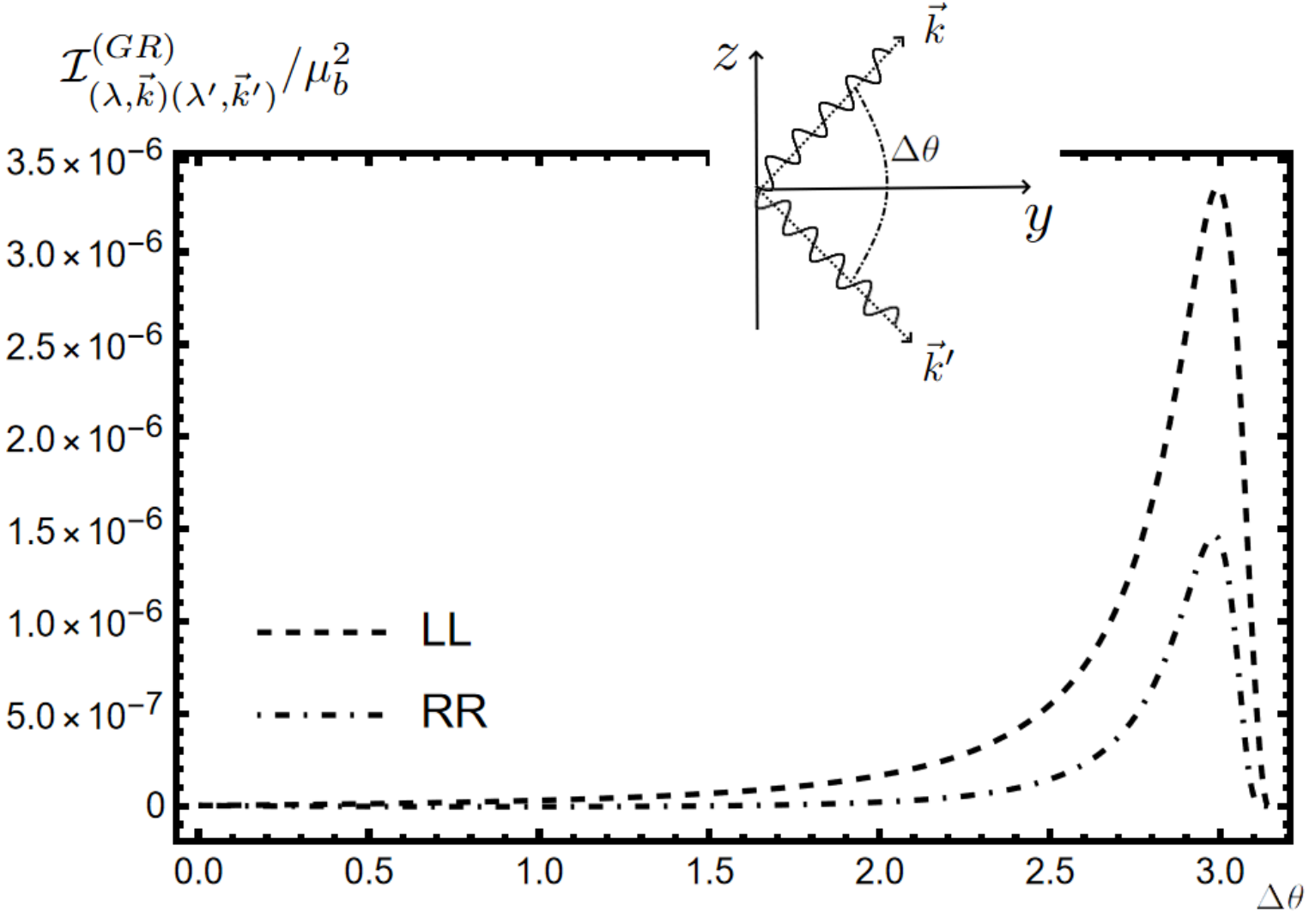}\hfill \includegraphics[width=0.9\linewidth]{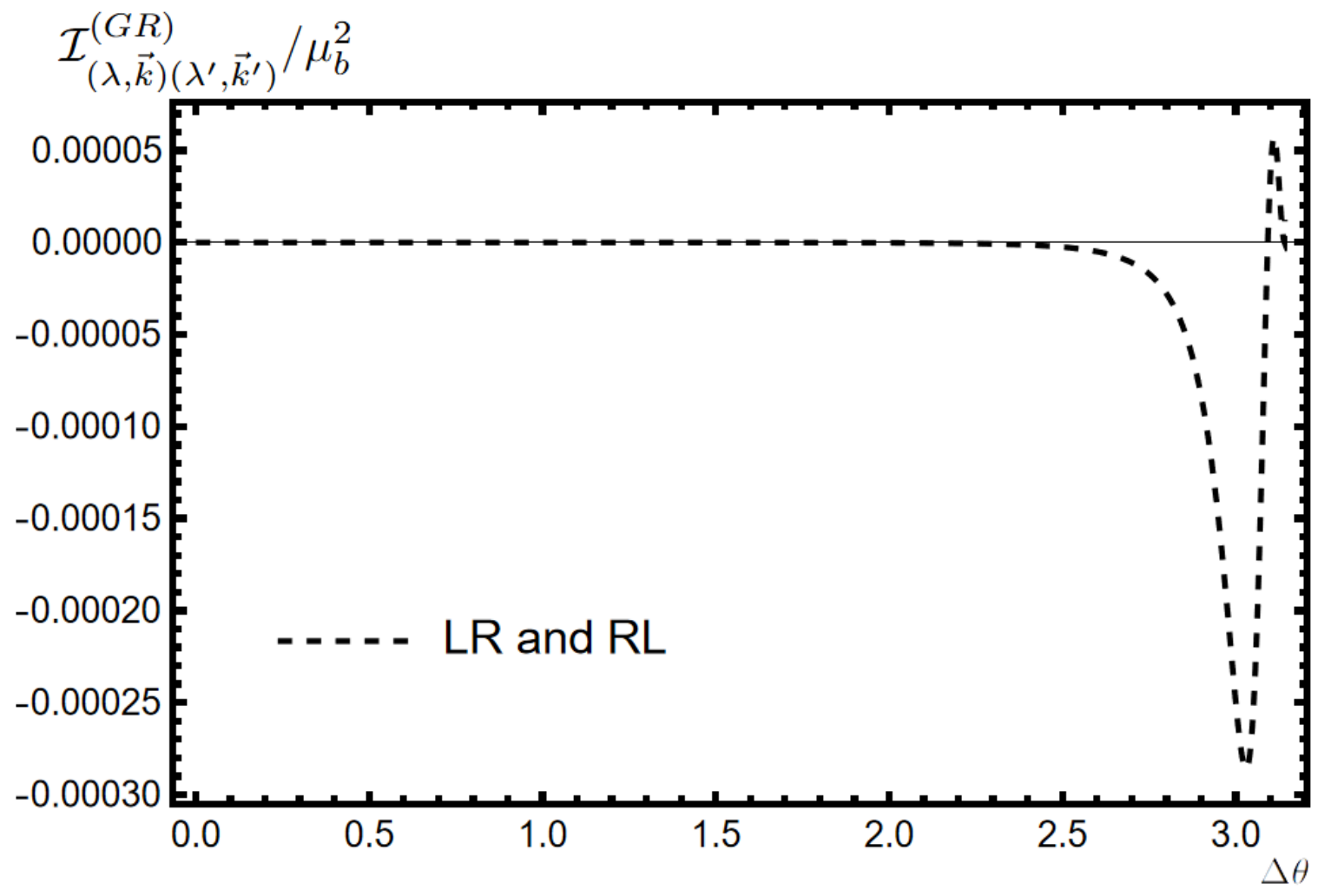}
    \caption{Angular and polarisation correlations for the GR interaction. The $2p$-state results in the asymmetry between the LL and RR pairs. The plot corresponds to $a_\mu=0.1$.}
    \label{GR_correlations}
\end{figure}

For the GR-induced interaction \eqref{S2}, we obtain the following expression,
\begin{equation}
\mathcal{F}^{(GR)}_{(\vec{k},\lambda)(\vec{k}^\prime,\lambda^\prime)} = \frac{  f_k f_{k^\prime} N_{2p}}{4\mu_b} \mathcal{I}^{(GR)}_{(\lambda,\vec{k})(\lambda^\prime,\vec{k}^\prime)} \  \ \ ,
\label{F_GR}
\end{equation}
where 
\begin{equation}
\label{I_GR_Correlation}
     \mathcal{I}^{(GR)}_{(\lambda,\vec{k})(\lambda^\prime,\vec{k}^\prime)}=e^{(\lambda)}_{im}(\vec{k}) I_{ij}(\vec{k},\vec{k}^\prime \ )e^{(\lambda^\prime)}_{mj}(\vec{k}^\prime)\,,
\end{equation}
and $I_{ij}$ has the tensorial structure  ,
\begin{equation}\label{Iij}
    I_{ij}(\vec{k},\vec{k}^\prime)=\int d^{3} \vec{x} \ \partial_i \Psi_{2p}(\vec{x})\partial_j \Psi_{2p}(\vec{x})e^{-i(\vec{k}+\vec{k}^\prime)\cdot \vec{x}}\, ,
\end{equation}
derived from the source. Since two axions are involved in the process, $E=2\mu_b$. The axion occupation number $N_{2p}$ acts as an enhancement parameter (in analogy to the quantum-optics SFWM). The angular and polarisation correlations of the gravitons emitted from the axionic cloud are determined from \eqref{I_GR_Correlation}. Considering gravitons emitted in the $y-z$ plane with $\vec{k}+\vec{k}^\prime$ corresponding to angles $\theta_{k+k^\prime}=\varphi_{k+k^\prime}=\pi/2$, we can express \eqref{I_GR_Correlation} in terms of the angle $\Delta\theta$ between the two gravitons ({\it c.f.} figure \ref{GR_correlations}). 

The axion decay process \eqref{S2CS} induced by the gravitational CS anomaly \eqref{Rcs}, yields:
\begin{equation}\label{IijFour}
   \mathcal{F}^{(CS)}_{(\lambda,\vec{k})(\lambda^\prime,\vec{k}^\prime)}=   iA\sqrt{\frac{N_{2p}}{2\mu_b}}f_k f_{k^\prime} \ \Omega^{2}_k 
 \ \Omega^{2}_{k^\prime}\ \mathcal{I}^{(CS)}_{(\lambda,\vec{k})(\lambda^\prime,\vec{k}^\prime)} \ ,
\end{equation}
where 
\begin{equation}
\label{I_CS_Correlations}
\begin{aligned}
    &\mathcal{I}^{(CS)}_{(\lambda,\vec{k})(\lambda^\prime,\vec{k}^\prime)}=l_{\vec{k}^\prime}l_{\lambda^\prime}\widetilde{\Psi}_{2p}(-\vec{k}-\vec{k}^{\prime})\times\\
    &\times\left(   [e^{(3)}(\vec{k}^\prime)] _m  \ e^{(\lambda)}_{mj}(\vec{k})\  e^{(\lambda^\prime)}_{jl}(\vec{k}^\prime) [e^{(3)}(\vec{k})] _l 
   \  +\right.\\
    &\left. + \left(1-   \cos\Delta\theta \right)  \ e^{(\lambda)}_{mj}(\vec{k}) \   e^{(\lambda^\prime)}_{mj}(\vec{k}^\prime)      \     \right) \ ,
    \end{aligned}
\end{equation}
with $e^{(3)}( \vec{k} ) =\vec{k}/| \vec{k} |$, and $\widetilde{\Psi}_{2p}(\vec{k})$ denotes the Fourier transform of $\Psi_{2p}(\vec{x})$, $l_\lambda=\pm1$, for $\lambda=L,R$ and $l_{\vec{k}}=\pm 1$, for $\theta_k>\pi/2$ or $\theta_k<\pi/2$, respectively, 
where $\theta_k$ is the polar angle of $\vec{k}$ \cite{Alexander:2004wk}.  Here  $E=\mu_b$ and the proportionality factor is $\sqrt{N_{2p}}$,  since  only one axion is involved in the microscopic process (as in SPDC).

\begin{figure}[ht]
    \centering
    \includegraphics[width=0.9\linewidth]{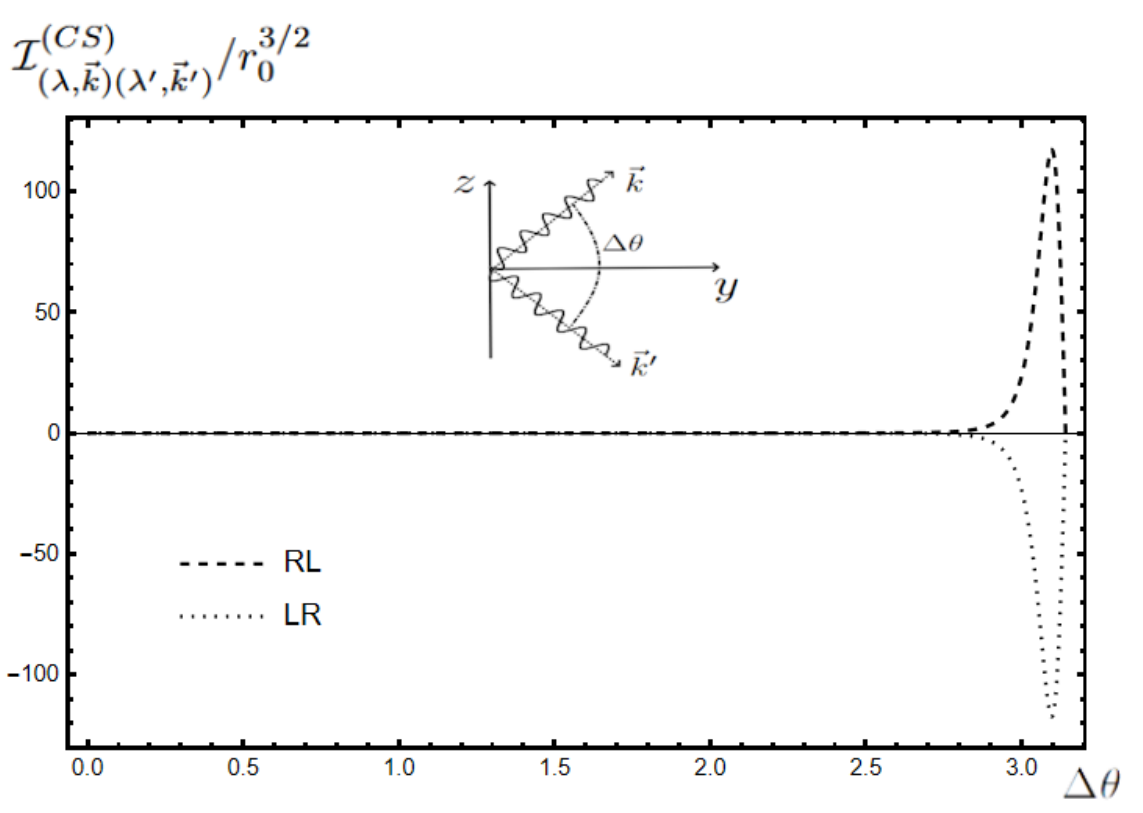}
    \caption{Angular and polarisation correlations for the CS interaction. Only pairs of opposite polarisations are produced; Maximal entanglement occurs between the L and R polarisations. The plot corresponds to $a_\mu=0.1$. }
    \label{CS_correlations}
\end{figure}

Both correlators \eqref{F_GR} and \eqref{IijFour}  are not separable states of two gravitons, indicating  that the graviton pairs are {\it entangled} \cite{Law:2000hyw}.  Due to the {\it non-relativistic regime} of the axionic cloud the graviton pairs are emitted approximately in opposite directions (i.e. {\it almost anti-collinear})  as shown in figures \ref{GR_correlations} and \ref{CS_correlations}.

{\it\textbf{ Multimode Squeezed State}}. Since the evolution operator \eqref{Evolution_Operator} has the form of a multimode squeezing operator, one can see that~\cite{multimode}: 
\begin{align}
\label{SaS_multimode}
    \hat{S^\dagger} \hat{\alpha}_{I} \hat{S} =\sum_J (\mu_{IJ}\hat{\alpha}_J  +  \nu_{IJ}\hat{\alpha}^\dagger_{J})\ , 
\end{align}
with the transformation coefficients given by:
\begin{align}
\label{mu_IJ_expansion} 
\mu_{IJ}=\delta_{IJ} + \frac{1}{2!}\sum_M \mathcal{G}_{IM}\mathcal{G}^\star_{MJ} \ + \ \dots \,,
\end{align}
\begin{align} 
\label{nu_IJ_expansion}
\nu_{IJ} = \mathcal{G}_{IJ} + \frac{1}{3!}\sum_{M,L} \mathcal{G}_{IM}\mathcal{G}^\star_{ML} \mathcal{G}_{LJ} \ + \ \dots \,  \ .
\end{align}
Using these transformations, one can estimate the average number of gravitons $N_{gr}$ in the squeezed vacuum state $|\psi\rangle=\hat{S}\vert 0 \rangle $:
\begin{align}\label{Nmax}
\langle N_{gr} \rangle = \sum_{I, J} 
\Big|\nu_{IJ}\Big|^2\ \lesssim \sum_{I,J} \Big( \Big|{\mathcal G}_{IJ}\Big|^2 \ + \  \dots \Big)\,,
\end{align}
where we have used \eqref{nu_IJ_expansion}, the $\dots$ denote the contributions from the cubic and higher-order terms in \eqref{nu_IJ_expansion}, 
and, for the upper bound, we use the triangle inequality.
In the case of a single-mode squeezed vacuum~\cite{Guerreiro:2019vbq}, $\mathcal{G}_{IJ}\sim\delta_{IJ}$, the expression reduces to the well known relation $\langle N_{gr}\rangle=\sinh^2 r$. From 
\eqref{nu_IJ_expansion}, \eqref{Nmax}, one observes that, in case $\sum_{I,J} \big|(\mathcal{G}_{IJ})\big|^{2}\ll1$,
the upper bound in \eqref{Nmax} is saturated,
implying that the average number of gravitons in the squeezed-vacuum state is highly suppressed. On the other hand, if the first term of the infinite series \eqref{Nmax} (or \eqref{nu_IJ_expansion}) is of order one or higher, then all terms in the series should be resummed, and the number of squeezed gravitons shows an exponential-like enhancement, similar to the single mode case.  
\par 
For the GR interaction, we consider the case of figure \ref{GR_correlations}.
Upon substituting \eqref{G_IJ} into \eqref{Nmax}, we find the following differential expression per solid angles
\begin{align}
    \nonumber 
    &\frac{d^2}{d\Omega d\Omega^\prime}\sum_{I,J} \Big|(\mathcal{G}^{(GR)}_{IJ})\Big|^2=\frac{1}{128\pi^3}\left(\frac{\mu_b}{M_{\rm Pl}}\right)^{4}N^{2}_{2p}\,T \mu_b \\ 
    &
\times\sum_{\lambda,\lambda^\prime}\int d\tilde{k}d\tilde{k}^\prime \ \tilde{k}\tilde{k}^\prime\delta\left(
    \tilde{k}+ \tilde{k}^\prime -2\right)\Big | \widetilde{\mathcal{I}}^{(GR)}_{(\lambda,\vec{k})(\lambda^\prime,\vec{k}^\prime)}\Big|^{2} \ ,
    \label{N_gr_per_Solid_angles}
\end{align}
where $\tilde{k}^{(\prime)}=k^{(\prime)}/\mu_b$ and $\mathcal{I}=\widetilde{\mathcal{I}}/\mu_b^2$. 
In order to estimate the number of gravitons in the squeezed vacuum state, we consider $\vec{q}=-(\vec{k}+\vec{k}^\prime)$ lying on $\theta_{k+k^\prime}=\varphi_{k+k^\prime}=\pi/2$ ({\it c.f.} figure \ref{GR_correlations}). The main contribution
to the integral \eqref{N_gr_per_Solid_angles} comes from momentum vectors $\vec k$, $\vec k^\prime$  satisfying $k\approx k^\prime\approx\mu_b$, whilst the maximum contribution corresponds to their relative angle $\Delta \theta=0.964\pi$. 
The integrations over the solid angles will yield 
a factor at most of 
order $\left(4 \pi \right)^2$, implying the following upper limit,

\begin{equation}
    \label{N_gr_RESULT}
\sum_{I,J}\vert\mathcal{G}^{(GR)}_{IJ}\vert^2 \lesssim 2.5\times10^{-15} \,T \mu_b\, .
\end{equation}
Note here that the large suppression induced by the ratio $\big(\mu_b/M_{\rm Pl}\big)$ is compensated by the number of axions in the cloud ,  
$ N_{2p}\left( \mu_b/M_{\rm Pl}\right)^2\approx 10^{-3}a_\mu^2 =10^{-5}\ 
$ (see \eqref{number_of_axions}).\par 

This is not the case for the anomaly induced interaction. In an effective field theory, the CS-coupling to the axion is suppressed below the cut-off scale of the theory.  
The pertinent squeezing effect is proportional to the coupling strength of the axion to the (gravitational) CS anomaly.
For the string-inspired KR axion coupling \cite{Duncan:1992vz} $A\sim 10^{-2}M_{\rm Pl}/M^{2}_\text{s}$, with $M_\text{s}$ the string scale.   As in the GR case, we argue ({\it c.f.} figure~\ref{CS_correlations}):
\begin{equation}
\label{N_cs_RESULT}
    \sum_{I,J} \Big|(\mathcal{G}^{(CS)}_{IJ})\Big|^2 \lesssim 10^{-10}\left(\frac{\mu_b}{M_s}\right)^4 \, \mu_b T\ .
\end{equation}

It is evident from \eqref{N_gr_RESULT} and \eqref{N_cs_RESULT} that the lifetime  $T$  of the axionic cloud is  also a  parameter able to induce significant squeezing. In contrast to the short lived quasi-normal modes (QNMs) \cite{BH_Squeezer}, the longevity of axionic clouds seems to overcome limitations and produce appreciable squeezing effects. We shall identify $T$ with the lifetime of the cloud, $T=\tau_{\text{cloud}}$. The lifetime of the axionic condensate is estimated in the literature by means of perturbative calculations and analytical approximations~\cite{BritoScales1,BritoScales2,Yoshino_2014_scales,porto_scales,BritoCardoso}, and all seem to agree that there is a clear separation of scales, i.e. $\tau_{\text{cloud}}\gg \tau_s$.  For example, in \cite{porto_scales}, the lifetime of the cloud results in a factor $T\sim 10^7 \tau_s$, for which, on using \eqref{timescale_superradiance} and  ~\eqref{N_gr_RESULT}, yields, 
\begin{equation}
    \label{N_gr_RESULT_final}
  \sum_{I,J}\vert\mathcal{G}^{(GR)}_{IJ}\vert^2 \lesssim 60 \   .
\end{equation} 
Hence, on account of \eqref{nu_IJ_expansion} and \eqref{Nmax}, 
the total number of gravitons in the squeezed vacuum state $N_{gr}$ 
is given by an infinite sum of positive integer powers of quantities of order given typically by \eqref{N_gr_RESULT_final}, which leads to an exponential enhancement 
of $N_{gr}$, 
resembling the $\langle N_{gr}\rangle= \sinh^{2}r$ behavior of the one-mode case \cite{Scully_Zubairy_1997}. From \eqref{N_gr_RESULT_final}, we obtain an estimate of the upper bound of the squeezing parameter 
in our multi-mode squeezed-graviton case
$r_{\rm multi-mode}^2 \equiv     \sum_{I,J}\vert\mathcal{G}^{(GR)}_{IJ}\vert^2 \lesssim 60$,
which would correspond to an upper bound on the average number of squeezed gravitons 
$\langle N_{gr}\rangle \lesssim \mathcal O(10^{7})$.
We note at this point that, depending on the detailed parameters of the BH and axion masses, we obtain a range $\sum_{I,J}\vert\mathcal{G}^{(GR)}_{IJ}\vert^2 = {\mathcal O}(10-60)$,
with the upper limit 
\eqref{N_gr_RESULT_final} being pretty robust in the non-relativistic case we consider here. In the relativistic case, where curvature effects cannot be avoided, the almost anti-collinear emission of graviton polarisation states breaks down, which might lead to additional enhancement of the squeezing. 
In contrast, the CS anomaly-induced squeezing \eqref{N_cs_RESULT} is still highly suppressed, compared to that induced by GR, since higher curvature interactions are subleading at low energy scales \cite{Donoghue:1994dn}, determined by the axion mass, $\mu_b$.  

 {\it\textbf{Observational Prospects and Outlook.}} Before closing we would like to make some remarks on the experimental observability of the multi-mode-graviton squeezing parameters associated with  \eqref{N_gr_RESULT_final}. As already mentioned, the non-observation of squeezed single-mode graviton states by LIGO/Virgo interferometers~\cite{LIGOScientific:2016aoc,McCuller:2021mbn} implies~\cite{Hertzberg:2021rbl} an upper bound on the pertinent squeezing parameter $r < 41$. The fact that a GW involves a continuum of modes, prompted the authors of \cite{Hertzberg:2021rbl} to consider (the more realistic case of) a Gaussian profile of squeezed modes, with the peak corresponding to the characteristic frequency of the classical GW as measured by the detector in a given event. This is how these authors derived their estimate on the upper bound of the allowed region of $r$. If these results are applied at face value to our case, then from \eqref{N_gr_RESULT} one can constrain the axionic-cloud life time, and thus falsify models with too long lifetimes, such as the one leading to the upper bound in \eqref{N_gr_RESULT_final}. 

We note, however, that an extension of the (Gaussian) wave-packet approach of \cite{Hertzberg:2021rbl} to our multi-mode  squeezed graviton states needs to be done before definite conclusions are reached on the allowed region of the pertinent squeezing parameters in our case, based on the LIGO/Virgo data. We also remark at this stage that an important feature of our suggested collective multi-mode graviton states is that they are 
{\it entangled} in polarisation. Because of this entanglement property, which is purely quantum in origin, the difficulties faced in the prospects of detecting single gravitons by simply looking for clicks in the detectors~\cite{singlegrav} might be avoided. Indeed, as discussed in \cite{Carney:2023nzz}, such ``click''-effects are mimicked by single-mode {\it classical} GW, which does not apply to our case. A combination of methods suggested in \cite{Carney:2023nzz}, in this respect, with potential, currently unachieved, but not impossible, graviton-polarisation and correlation statistics measurements, might be a way forward. The above constitute important but challenging future research avenues which fall beyond the scope of our analysis here. 
We hope, nonetheless, that our work provides a motivation for experimental searches and theoretical feasibility studies of such phenomena in future interferometers. Given the current LIGO/Virgo sensitivities~\cite{LIGOScientific:2016aoc,McCuller:2021mbn,Hertzberg:2021rbl}, 
such future instruments will have the sensitivity  to falsify our models, via exclusion of too-long axion lifetimes.

Our analysis above has been  restricted to the flat-background approximation and crucially depends on the longevity of the axionic clouds. 
Extending the situations to curved spacetimes in the neighborhood of the Kerr solution is a significant technical and conceptual challenge, which we did not discuss here. In early-Universe cosmology (inflation), spacetime curvature effects are known to lead to enhancement of the graviton squeezing. Whether such a situation persists in our local case is not known. Nonetheless we refer the reader to \cite{Chatzifotis:2022mob,Chatzifotis:2022ene}, where back reaction effects of axions on the Kerr geometry in the context of CS gravity have been studied in conjunction with the strength of the CS coupling, which above a critical value leads to significant effects on the BH angular momentum. It would also be interesting but challenging to extend the approach  to take into account axion self interactions. All such effects may provide additional modifications (perhaps enhancement) on graviton squeezing.

\color{black}
\begin{center}
  \line(1,0){70}
 \end{center}

{\it Acknowledgments} The authors would like to thank the anonymous referees for their constructive feedback. The work of P.D. is supported by a graduate scholarship from the National Technical University of Athens (Greece).
The work of NEM and SS is supported in part by the UK Science and Technology Facilities research Council (STFC) under the research grant  ST/X000753/1,  and by the UK Engineering and Physical Sciences Research Council (EPSRC) under the reserach grant No. EP/V002821/1. 
The work of S.-N.V. is supported by the Hellenic Foundation for Research and Innovation
(H.F.R.I. (EL.ID.EK.)) under the “5th Call for H.F.R.I. Scholarships to PhD Candidates” (Scholarship Number:
20572). NEM also acknowledges participation in the COST Association Actions CA21136 “Addressing observational
tensions in cosmology with systematics and fundamental physics (CosmoVerse)” and CA23130 ``Bridging high and
low energies in search of quantum gravity (BridgeQG)”.

\bibliographystyle{apsrev4-2}
\bibliography{squeezing}

\end{document}